\documentclass[final]{svjour3}
\usepackage{graphicx}
\usepackage{rotating}
\usepackage{amssymb}
\usepackage{mathptmx}
\usepackage[numbers]{natbib}
\usepackage{xcolor}
\makeatletter
\journalname{Journal of Low Temperature Physics}
\newcommand{\orcid}[1]{\href{https://orcid.org/#1}{\textcolor[HTML]{A6CE39}{\aiOrcid}}}
\usepackage{tikz,xcolor,hyperref}
\usepackage{ulem}

\definecolor{lime}{HTML}{A6CE39}
\DeclareRobustCommand{\orcidicon}{%
	\begin{tikzpicture}
	\draw[lime, fill=lime] (0,0) 
	circle [radius=0.16] 
	node[white] {{\fontfamily{qag}\selectfont \tiny ID}};
	\draw[white, fill=white] (-0.0625,0.095) 
	circle [radius=0.007];
	\end{tikzpicture}
	\hspace{-2mm}
}

\foreach \x in {A, ..., Z}{%
	\expandafter\xdef\csname orcid\x\endcsname{\noexpand\href{https://orcid.org/\csname orcidauthor\x\endcsname}{\noexpand\orcidicon}}
}

\begin{document}

\newcommand{\hdblarrow}{H\makebox[0.9ex][l]{$\downdownarrows$}-}
\title{Investigation of Optical Coupling in Microwave Kinetic Inductance Detectors using Superconducting Reflective Plates}

\author{Paul Nicaise\textsuperscript{1}\orcidA{}, Jie Hu\textsuperscript{1, 2}\orcidC{}, Jean-Marc Martin\textsuperscript{1}\orcidD{}, Samir Beldi\textsuperscript{3}, Christine Chaumont\textsuperscript{1}, Piercarlo Bonifacio\textsuperscript{1}, Michel Piat\textsuperscript{2}, Hervé Geoffray\textsuperscript{4} and Faouzi Boussaha\textsuperscript{1}\orcidB{}}

\institute{1. GEPI, Observatoire de Paris, PSL Université, CNRS, 75014 Paris, France \\
2. APC, Université de Paris, CNRS, 75013 Paris, France \\
3. ESME Research Lab, 94200 Ivry-sur-Seine - France \\
4. CNES, 31400 Toulouse, France \\
\email{paul.nicaise@obspm.fr}}

\maketitle

\begin{abstract}

To improve the optical coupling in Microwave Kinetic Inductance Detectors (MKIDs), we investigate the use of a reflective plate beneath the meandered absorber. We designed, fabricated and characterized high-Q factors TiN-based MKIDs on sapphire operating at optical wavelengths with a Au/Nb reflective thin bilayer below the meander. The reflector is set at a quarter-wave distance from the meander using a transparent Al$_2$O$_3$ dielectric layer to reach the peak photon absorption. We expect the plate to recover undetected photons by reflecting them back onto the absorber.


\end{abstract}

\section{Introduction}

Over the past decade~\cite{meeker18,walter20}, the study of Microwave Kinetic Inductance Detectors (MKIDs) in the optical to near-infrared range has been growing exponentially to take advantage of their various benefits such as single photon counting, intrinsic energy resolution and scalability into large array, all of which make optical MKIDs very attractive for ground-based astronomical applications that images faint objects. This is precisely the case for the SpectroPhotometric Imaging in Astronomy with Kinetic Inductance Detectors (SPIAKID) project. We aim to build a broadband 0.4-1.6~µm MKID-based camera that will be housed in the 3.58-meter New Technology Telescope (NTT) in Chili in 2025 to derive the age and metallicity from stars that compose Ultra-Faint Dwarf galaxies in the Local Group.
This challenging goal requires improvement regarding the relatively low optical coupling efficiency of MKIDs that does not exceed 30\% depending on the material used~\cite{szypryt16,mazin13}.
In recent years, considerable efforts have been made to increase absorption by using anti-reflection coatings above the absorber part of the pixel that shows promising results~\cite{mccarrick13,dai19}.
However, a significant amount of photons in the visible range are lost by transmission through the absorber layer.
This is especially true as we try to increase the overall pixel sensitivity by reducing the volume of the absorber and hence its thickness~\cite{guo17}.
Here, we propose to add a reflecting layer at a quarter-wave distance below the absorber to retrieve photons that have been transmitted through. 

\section{Design parameters}\label{sec:param}

The resonator consists of a 60~nm stoechiometric titanium nitride (TiN) layer with a critical temperature $T_c=4.6~K$ which corresponds to a kinetic inductance per square $L_{k,TiN} = 7~pH/ \square$. It has a 42$\times$37~µm$^2$ meandered inductor with 0.7~µm gaps and a large 400$\times$400~µm$^2$ Interdigitated Capacitor (IDC) with 1~µm wide fingers and 1~µm gaps. The resonator is capacitively coupled to a Coplanar Waveguide (CPW) feedline made of a 100~nm niobium layer with a kinetic inductance per square $L_{k,Nb} = 0.2~pH/\square$.
These superconducting layers are deposited on a 280~µm thick sapphire substrate where Two-Level System (TLS) noise is less prominent than in amorphous materials.
The CPW central strip is 5~µm wide with 2~µm gaps to ensure the characteristic impedance to be $Z_0=50\ \Omega$.

\begin{figure}[htbp]
\begin{center}
\includegraphics[width=0.6\linewidth, keepaspectratio]{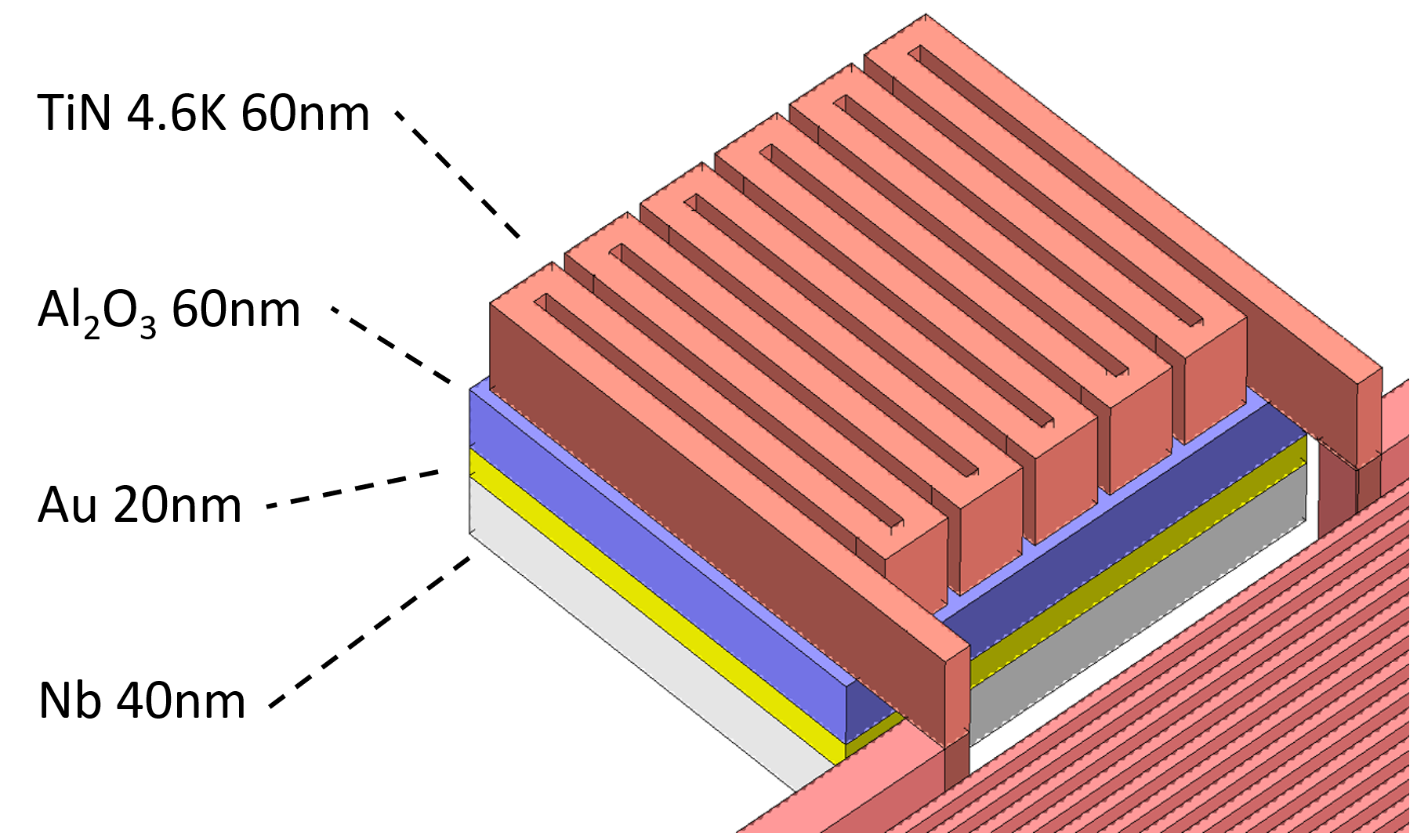}
\caption{3D geometry of the simulated TiN inductor and optical stack. The TiN meandered absorber rests on the optical stack that consists from top to bottom of: an aluminium oxide spacer layer, a gold reflective layer and a niobium layer to minimize electrical loss in the non-superconductive gold layer through the proximity effect.} \label{fig:pixel}
\end{center}
\end{figure}

\begin{figure}[htbp]
\begin{center}
\includegraphics[width=1\linewidth, keepaspectratio]{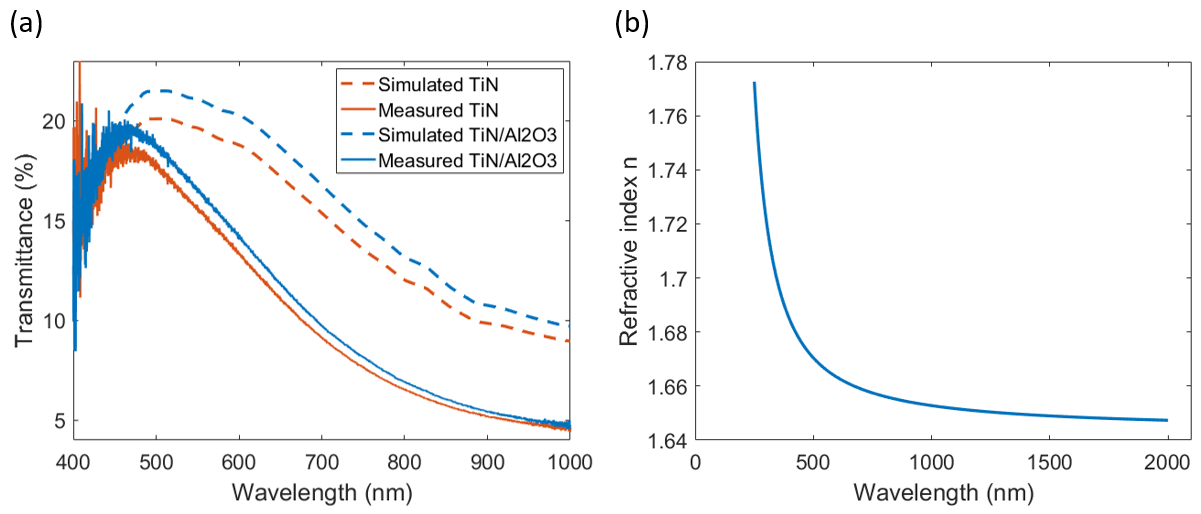}
\caption{(a) Simulated and measured transmission at room temperature in red of a 60nm TiN 4.6K layer and in blue with the addition of a 60nm Al$_2$O$_3$ spacer layer below the TiN. In both the simulations and measurements, the spacer layer slightly increases transmittance. These measurements have been made on uniform films. For a TiN layer patterned as a meander, the transmittance could be increased. (b) Refractive index n of the Al$_2$O$_3$ layer measured by ellipsometry.} \label{fig:transmittance}
\end{center}
\end{figure}

As presented in Fig~\ref{fig:pixel}, there are multiple layers beneath the meander that make up the optical stack. First is an aluminium oxide (Al$_2$O$_3$) insulating layer that separates the absorber from the reflecting layer. Al$_2$O$_3$ can feature a very low dielectric loss tangent around $\mathrm{tan} \delta \approx 10^{-6}$ by using the Atomic Layer Deposition (ALD) process~\cite{beldi19}. Furthermore, ALD allows to reach excellent film uniformity compared to other deposition methods.
In order to assess the transmission of the sputtered TiN layer and Al$_2$O$_3$ layer deposited by ALD, we performed spectrophotometry measurements. As shown in Fig~\ref{fig:transmittance}(a), the maximum transmission of TiN in the visible range occurs at around $\lambda_{max}$ = 450 nm as also reported elsewhere~\cite{dai19}.
According to the measured refractive index obtained by ellipsometry and shown in Fig~\ref{fig:transmittance}(b), the corresponding index is $n= 1.675$. To satisfy the quarter-wave condition, this leads to an optimal Al$_2$O$_3$ thickness of $t_{Al_2O_3} = \frac{\lambda_{max}}{4n} \approx 67$ nm. Due to some constraints at the time regarding the deposition time of our ALD system, the thickness of the deposited  Al$_2$O$_3$ layer was measured to be around 60 nm, which is still close to the expected optimal thickness.
The blue curve in Fig~\ref{fig:transmittance}(a) confirms that adding the 60nm thick Al$_2$O$_3$ layer below the TiN layer does not deteriorate the transmission and even slightly improves it. This could be explained by a better impedance matching between the layers that reduces reflection which is supported by the dotted curves corresponding to Zemax simulations.

Below the spacer layer is the 20~nm gold layer that provides close to unity reflectance in the visible to near-infrared range. Gold is not a superconductor and still has a residual resistivity at sub-K temperatures. This leads to electrical losses that can deteriorate the internal quality factor of the resonator. To overcome this issue, we place a niobium layer  directly underneath so that the superconducting state of niobium is carried over to the gold layer thanks to the proximity effect. To maximize this effect, the thickness of the niobium layer must be at least equal to its coherence length $\xi_{Nb}\approx 40$~nm.

\section{Resonator simulations results}\label{sec:simulations}

\begin{figure}[htbp]
\begin{center}
\includegraphics[width=0.95\linewidth, keepaspectratio]{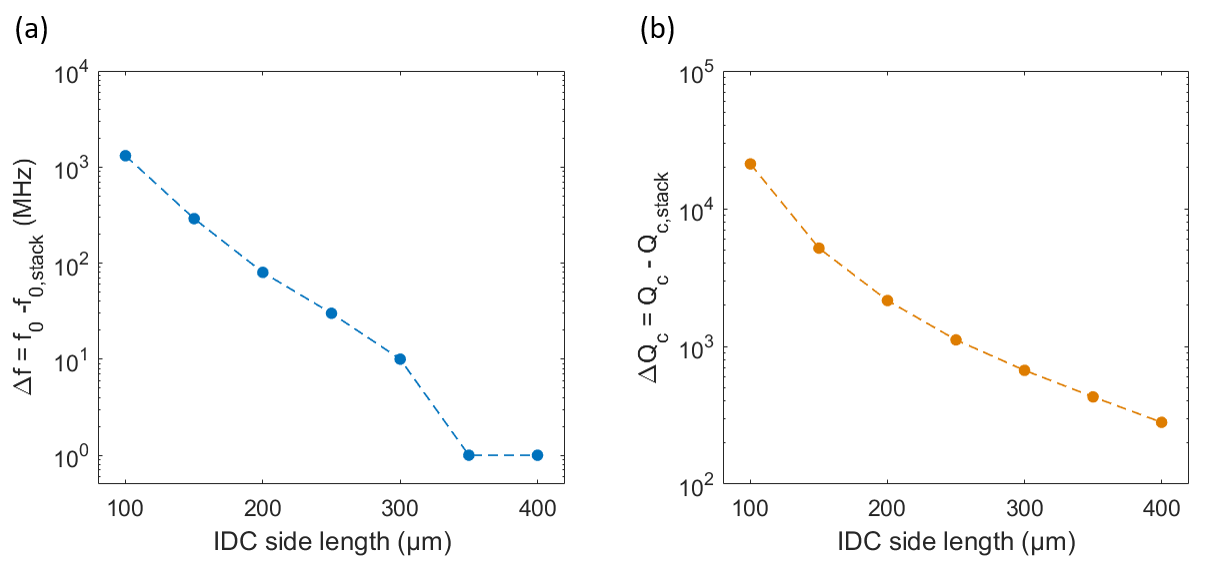}
\caption{Evolution of the resonator parameters with and without the optical stack as a function of IDC size.
(a) The frequency change $\Delta f_0 = f_0 - f_{0,stack}$ is no longer affected by the addition of the optical stack above an IDC size of 300x300 µm$^2$.
(b) The coupling quality factor change $\Delta Q_c = Q_c - Q_{c,stack}$ also decreases with larger IDC size.} \label{fig:idcsize_vs_dqc_df0}
\end{center}
\end{figure}

The whole LC structure, simulated on Sonnet, resonates at $f_0=1.41$~GHz with a coupling quality factor of $Q_c$ = 22600.
If we simulate the resonator with the optical stack, the resonance frequency is lowered to $f_{0,stack}=1.40$~GHz and $Q_{c,stack} = 22900$ which shows that the optical stack has almost no influence on the LC structure. However, this is not the case for a smaller 100$\times$100~µm$^2$ IDC where the optical stack lowers the frequency from 8.20~GHz to 6.84~GHz and increases the coupling quality factor from 16500 to 45600. Figs.~\ref{fig:idcsize_vs_dqc_df0}a and ~\ref{fig:idcsize_vs_dqc_df0}b clearly show that the optical stack has more influence on the resonance structure as the size of the IDC decreases.
This is certainly explained by a parasitic capacitor that arises from the optical stack. Indeed, the reflector, dielectric and meander form a Parallel-Plate Capacitor (PPC). Moreover, this may also introduce TLS noise which is why the choice of dielectric for the spacer layer is critical.
We can estimate the capacitance of the parasitic PPC on Sonnet using a two-port network.
We can then extract the admittance parameters and find the capacitance value using Eq.~\ref{eq:c}~\cite{sonnet}.
\begin{equation}
C(f) = \frac{10^{-12}}{2\pi f \times Im(1/Y_{21})}[pF]
\label{eq:c}
\end{equation}
where $f$ is the microwave frequency and $Y_{21}$ is the forward transfer admittance.\\

In our case, for $t=60$~nm, the PPC capacitance $C_{PPC}$ of $0.13$~pF does not contribute significantly to the total capacitance $C_{tot}=C_{IDC}+C_{PPC}=3.77$~pF. But for MKIDs with a smaller IDCs, the parasitic PPC can represent a significant amount of the total capacitance. For this reason, we decided to use the 400$\times$400~µm$^2$ to keep better control of the design parameters.

\section{Preliminary characterizations}

We fabricated several batches of 20-MKID arrays in the Paris Observatory cleanroom facility. The earliest arrays had similar designs to the one detailed in Sec.~\ref{sec:param} with a few exceptions. The first one had no niobium below the gold reflector and resulted in shallow resonances due to the lossy gold layer. The second one had the Au/Nb bilayer but the spacer layer used was silicon monoxide (SiO) that was chosen for the ease of deposition. The resonances were deeper thanks to the niobium layer but saturated at a fairly low readout power ($\leq$ -90~dBm). For this reason, we replaced SiO with Al$_2$O$_3$ to benefit from its much lower dielectric loss tangent.

\begin{figure}[htbp]
\begin{center}
\includegraphics[width=0.6\linewidth, keepaspectratio]{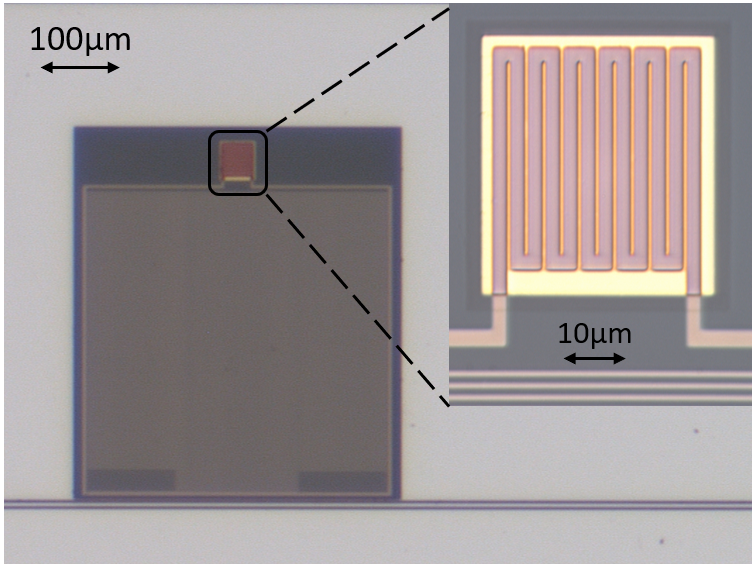}
\caption{Single MKID with a close-up view of the TiN inductor deposited on top of the optical stack. The yellow square represents the reflector and the slightly larger transparent square is the spacer layer that completely covers the edges of the reflector layer to avoid any short when depositing the resonator on top of it.} \label{fig:fabrication}
\end{center}
\end{figure}

\begin{figure}[htbp]
\begin{center}
\includegraphics[width=0.8\linewidth, keepaspectratio]{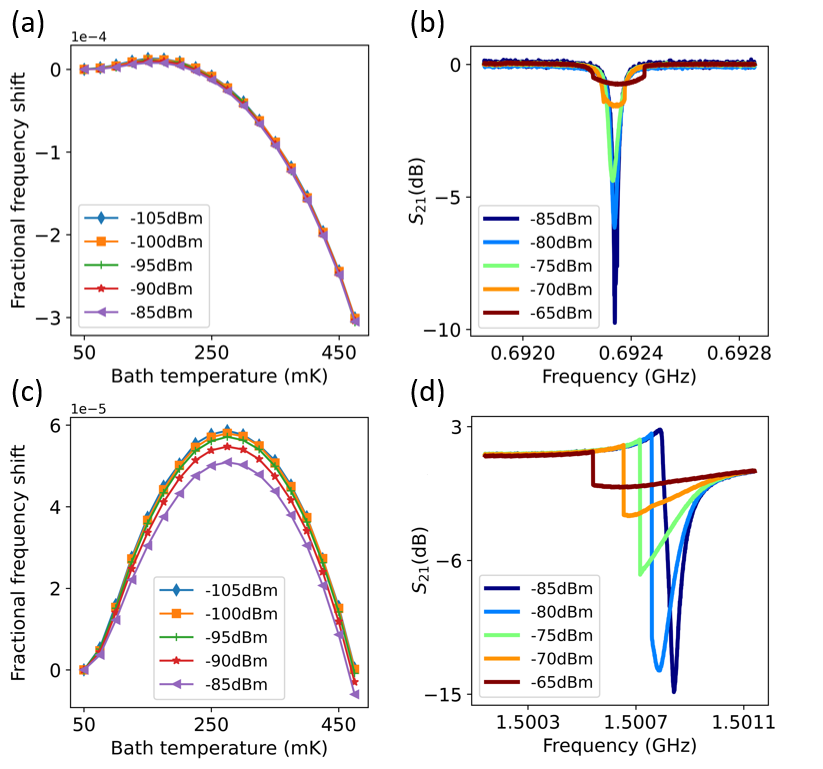}
\caption{Fractional frequency shift of (a) a MKID without the optical stack and (c) with the optical stack as a function of bath temperature. The different evolution from 50mK to 250mK between (a) and (c) highlights the signature of TLS noise in the MKID with the optical stack. The transmission at different readout power also reveals the influence of the optical stack on the resonator in (d) as it saturates and shift to lower frequencies for higher input powers compared to the pixel without reflector in (b).} \label{fig:measurements}
\end{center}
\end{figure}

For the design detailed in Sec.~\ref{sec:param}, the different layers were sputtered on a 2-inch sapphire wafer except for the Al$_2$O$_3$ spacer layer that was deposited using ALD and the gold layer that was deposited using thermal evaporation. The lithography and deposition parameters are identical to S. Beldi's process~\cite{beldi19}. This first batch consists of 10 resonators with an optical stack like the one showed in Fig.~\ref{fig:fabrication} and 10 without.
The wafer is wire-bonded to two 50~$\Omega$ SMA connectors in a gold-plated copper box and is cooled down in an Adiabatic Demagnetization Refrigerator (ADR) at APC~\cite{Jie20}. The first characterization were done in the dark at temperatures ranging from 50~mK to 500~mK and readout powers ranging from -105~dBm to -65~dBm.
Fig.~\ref{fig:measurements} highlights the increase of TLS noise in a MKID with the optical stack. Indeed, as the bath temperature rises from 50~mK to 250~mK in Fig.~\ref{fig:measurements}c, the resonance first shifts to higher frequencies which is the typical signature of TLS~\cite{wheeler20}. Then, from 250~mK upwards, the Cooper pair population starts decreasing and thus the resonance shifts back to lower frequencies as the kinetic inductance increases. This signature of TLS is not present for the pixel without the optical stack in Fig.~\ref{fig:measurements}a where the fractional frequency shift stays almost constant between 50~mK and 200~mK and then rapidly goes to lower frequencies because of the expected increase in kinetic inductance. The same conclusion can be drawn from Fig.~\ref{fig:measurements}d where the resonance at 50~mK starts to saturate from -80~dBm upwards. For Fig.~\ref{fig:measurements}b, the resonance gets shallower with increased readout power but however does not saturate.

\section{Conclusion and further investigations}

The addition of an optical stack below the inductive absorber creates a parasitic capacitance that affects the resonance frequency and quality factors. This phenomenon is amplified for smaller IDC sizes that have a capacitance of the same order as the PPC. The choice of aluminium oxide for the spacer layer is important to minimize dielectric loss and reach high quality factors. A gold reflector is used to reach unity reflection in the 0.5-1.6~µm range but it needs to be coupled with a superconductor like niobium to reduce electrical losses.
The preliminary characterizations show the signature of TLS noise for a MKID with the optical stack. The next step will be to measure the pulse response of this array at different wavelengths and further investigate the TLS effects in reflector-based optical MKIDs.

\begin{acknowledgements}
We thank Pierre Baudoz and Aurélien Pelleau at LESIA for the thin layer transmission measurements as well as Alexine Marret at GEPI for the Zemax simulations. Thank you to Florent Reix and Josiane Firminy at GEPI for wafer cutting and wire bonding. We also thank Michael Rosticher at ENS for aluminium oxide deposition and Xavier Lafosse and Lina Gatilova at C2N for the refractive index measurements. This work is supported by the European Research Council (ERC) and Centre National d'Etudes Spatiales (CNES).
\end{acknowledgements}

\pagebreak

\end{document}